\documentclass[a4paper,12pt]{article}

\usepackage{amsmath}
\usepackage{amssymb,epsfig,psfig}
\usepackage{amsfonts}
\usepackage{graphicx}
\usepackage{float}
\usepackage{latexsym}    
\newtheorem{lemma}{Lemma}

\def\ZZ{\hbox to 8.2222pt{\rm Z\hskip-4pt \rm Z}}

\newcommand{\de}{\delta}

\newcommand{\la}{\lambda}

\newcommand{\Om}{\Omega}
\newcommand{\Si}{\Sigma}

\newcommand{\Ga}{\Gamma}

\newcommand{\be}{\begin{equation}}
\newcommand{\ee}{\end{equation}}
\newcommand{\bqa}{\begin{eqnarray}}
\newcommand{\eqa}{\end{eqnarray}}
\newcommand{\ba}{\begin{array}}
\newcommand{\ea}{\end{array}}

\newcommand{\cS}{{\cal S}}
\newcommand{\cR}{{\cal R}}

\newcommand{\N}{\mathbb{N}}
\newcommand{\resetequ}{\setcounter{equation}{0}}

\begin{document} 

\title{Two and Three Loops Beta Function\\ of Non Commutative
 $\Phi^4_4$ Theory\footnote{Work supported by ANR grant NT05-3-43374 ``GenoPhy".}}
\author{Margherita Disertori$^1$\footnote{e-mail: Margherita.Disertori@univ-rouen.fr}, 
Vincent Rivasseau$^2$\footnote{e-mail: Vincent.Rivasseau@th.u-psud.fr}\\
1) Laboratoire de Math\'ematiques Rapha\"el Salem, UMR CNRS 6085\\
Universit\'e de Rouen, 76801\\
2) Laboratoire de Physique Th\'eorique, UMR CNRS 8627,\\
Universit\'e Paris-Sud XI, 91405 Orsay\\
}
\maketitle


\maketitle
 
\begin{abstract}
The simplest non commutative renormalizable field theory, the $\phi_4^4$ 
model on four dimensional Moyal space with harmonic potential 
is asymptotically safe at one loop, as shown by H. Grosse and R. Wulkenhaar. 
We extend this result up to three loops. If this remains true at any loop, it should allow a full non perturbative construction of this model.
\end{abstract}

\section{Introduction} 

Non commutative (NC) quantum field theory (QFT) may be important for physics 
beyond the standard model and for understanding the quantum
Hall effect \cite{DN}.
It also occurs naturally as an effective regime of string theory \cite{CDS} \cite{SW}.

The simplest NC field theory is the $\phi_4^4$ 
model on the Moyal space. Its perturbative renormalizability 
at all orders has been proved by 
Grosse, Wulkenhaar and followers \cite{GW1}\cite{GW2}\cite{RVW}\cite{GMRV}. 
Grosse and Wulkenhaar solved the difficult problem of ultraviolet/infrared
mixing by introducing a new harmonic potential term 
inspired by the Langmann-Szabo 
duality \cite{LS} between positions and momenta. 

Other renormalizable models of the same kind, including the orientable
Fermionic Gross-Neveu model have been recently also shown renormalizable at all orders \cite{V}, and techniques such as the parametric representation have
been extended to NCQFT \cite{GR}.
In view of these progresses it is tempting to conjecture that
commutative renormalizable theories in general have NC renormalizable
extensions to Moyal spaces which imply new parameters. However
the most interesting case, namely the one of gauge theories, still remains elusive
in this respect.

Returning to the NC $\phi_4^4$ theory, the next obvious step is the computation of the renormalization group
(RG) flow.
It is well known that the ordinary stable commutative $\phi_4^4$
model is not asymptotically free in the 
ultraviolet regime. The coupling is screened at lower momentum scales, or
conversely the bare coupling corresponding to a fixed (small) renormalized coupling
seems to explode as the cutoff is removed. This phenomenon is called the 
Landau ghost and should not be underestimated: it almost killed quantum field theory in the 60's! Field theory was resurrected by the discovery of ultraviolet asymptotic freedom in non-Abelian gauge theory in the early 70's, but the crisis left
unexpected byproducts. The main one is certainly
the accidental discovery of string theory itself, which evolved 
out of the Veneziano formula as an attempt
to bypass field theory in dual models.

An amazing discovery was made in \cite{GWbeta}:
the non commutative $\phi_4^4$ model does not exhibit any Landau ghost 
at one loop. It is not asymptotically free either: the RG flow
is simply bounded. The flow of the coupling goes from a small 
renormalized value to a larger but finite bare one. The difference 
increases when the Grosse-Wulkenhaar harmonic potential parameter $\Omega$ goes to 0.

Which gun killed the Landau ghost? NC $\phi_4^4$ has the same positivity
and stability as the commutative version, so the ``bubble graph"
must have the standard sign. It cannot vanish. The ``smoking gun"
is the wave function renormalization. We know that to measure the physical coupling 
requires the correct normalization of the four external fields, which in turn depends 
of the wave function renormalization. At one loop and in commutative $\phi_4^4$ field 
theory this wave function renormalization vanishes because the
``tadpole" graph is local. But it is no longer local in the NC $\phi_4^4$ model!
In general when the Grosse-Wulkenhaar parameter $\Omega$, which lies 
in $]0,1]$ is strictly smaller than 1, the beta function remains of 
the ordinary sign. But  at the special LS dual point $\Omega =1$ it vanishes.
With hindsight we may have predicted this
phenomenon because at $\Omega =1$ positions and momenta become
indistinguishable. Hence the flow should no longer distinguish where is 
the ultraviolet and infrared directions,
so that the coupling which no longer knows whether to grow or 
shrink, should remain constant...
Now the true marvel is that the flow of $\Omega$ itself always goes very fast to 
$\Omega =1$ in the ultraviolet. Therefore it blocks the growth of the
bare coupling and kills the Landau ghost!

This beautiful scenario is established in \cite{GWbeta} only at one loop.
In this paper we accomplish a new step to confirm it.
We compute the flow up to three loops at the special LS dual point $\Omega =1$,
and check that the beta function still vanish up to this order.
Equivalently up to three loops the
difference between bare and renormalized coupling remains 
finite. We establish this fact by brute force study of 
all planar four and two point graphs up to three loops. We need to take carefully
into account combinatoric factors, mass renormalization and loop symmetrization.
We obviously conjecture that the beta function vanishes to any order, 
but we have not been able to find the general proof yet.

The non perturbative construction of the model might follow
from our conjecture and a standard multiscale analysis. But some obstacles still 
remain on the road. One should for instance be able 
first to prove uniform Borel summability of the model 
in a single renormalization group slice (with slice-independent radius) 
through some kind of cluster-Mayer expansion \cite{GJ}\cite{R}.
This does not look easy because standard constructive techniques such as cluster 
expansions typically fail for large-matrix models. So we must warn the reader that there 
lies some exciting difficult work ahead before reaching the historic ``Graal" 
of constructive field theory, a full $\phi_4^4$ construction 
(on the unexpected non commutative Moyal space!).

\section{Notations and Main Result}
\resetequ

We follow the notations of \cite{GWbeta}. The propagator in the matrix base at $\Omega=1$ is
\be \label{propafixed}
C_{mn;kl} = \tfrac{1}{(4\pi)^2\theta} G_{mn} \delta_{ml}\delta_{nk} \ ; \ 
G_{mn}= \frac{1}{A+m+n}\  ,
\ee
where $A= 2+ \mu^2 \theta/4$, $m,n\in \mathbb{N}^2$ ($\mu$ being the mass)
and we used the abbreviations
\be
\de_{ml} = \de_{m_1l_1} \de_{m_2l_2}\ , \qquad m+m = m_1 + m_2 + n_1 + n_2 \ .
\ee

The vertex  for $\Phi^{*4}_4$ is:
\bqa
V_r = \frac{\lambda}{4}\ (4\pi^2 \theta^2)\  
\sum_{m,n,k,l\in \N^2} \phi_{mn} \phi_{nk} \phi_{kl} \phi_{lm} \ ,
\eqa
and for the $(\bar\Phi\star\Phi)^{*2}$ interaction of \cite{LSZ} it is
\be
V_c = \frac{\lambda}{2}\ (4\pi^2 \theta^2)\ \sum_{m,n,k,l\in \N^2}
 \bar\phi_{mn} \phi_{nk} \bar\phi_{kl} \phi_{lm} ,
\ee
so that the action is
\be
S_r =  \frac{(4\pi)^2 \theta}{2} \sum_{m,n\in\N^2}\ \phi_{mn} G^{-1}_{mn}  
\phi_{nm} +
\ \frac{\lambda}{4} (4\pi^2 \theta^2)\ \sum_{m,n,k,l\in \N^2}
\phi_{mn} \phi_{nk} \phi_{kl} \phi_{lm} 
\ee
or
\be
S_c = (4\pi)^2 \theta \sum_{m,n\in\N^2} \bar\phi_{mn} G^{-1}_{mn} \phi_{nm} +
\frac{\lambda}{2}\ (4\pi^2 \theta^2)\    \sum_{m,n,k,l\in \N^2} 
\bar\phi_{mn} \phi_{nk} \bar\phi_{kl} \phi_{lm} \ .
\ee
We have to compute the evolution equation of the effective coupling
\be
\la_r = -\frac{1}{4\pi^2\theta^2} \frac{\Ga_4(0,0,0,0)}{Z^2}
\ee
where the wave function normalization is
\be
Z = 1 -\frac{1}{(4\pi)^2 \theta} \partial_{m_1}\Si(m,n) \vert_{m,n=0}\ ,
\ee
with self energy
\be
\Si(m,n) = \langle \phi_{mn} \phi_{nm} \rangle_{1PI}^t \ .
\ee
The derivation is indeed equivalent to the difference definition of Grosse-Wulkenhaar.
The four point 1PI function is 
\be
\Ga_4(m,n,k,l) = \langle \phi_{mn} \phi_{nk} \phi_{kl} \phi_{lm} 
\rangle_{1PI}^t \ .
\ee
When computing Feynman graphs we must remember that
\begin{itemize}
\item each line comes with a factor 
$\frac{1}{(4\pi)^2\theta}$,
\item each vertex brings a factor $4\pi^2\theta^2\la/4$
for the real case and $4\pi^2\theta^2\la/2$ for the complex one.
\end{itemize}

\subsection{Main Result}
It is convenient to define $\tilde \lambda =\lambda/16\pi^2$,
$\lambda$ being the bare coupling. We now keep this bare coupling fixed. Our result 
states that the renormalized coupling
is a finite function of the bare coupling up to order 3
(in the limit where the ultraviolet cutoff goes to infinity):
\paragraph{Theorem}{\it 
At $\Omega=1$ we have
\be\label{mainresult}
\la_r = -\frac{1}{4\pi^2\theta^2} \frac{\Ga_4(0,0,0,0)}{Z^2} =  1 +  
\gamma_2\tilde\la^2- \gamma_3 \tilde\la^3 + O(\la^4)\ ,
\ee
where $\gamma_2$ and $\gamma_3$ are finite when the ultraviolet cutoff is removed\footnote{We could have included a $\gamma_1 \tilde\la$ term, but it turns out to be exactly zero.}.}

\vskip.5cm

We conjecture that this result holds up to any number of loops:
\paragraph{Conjecture: "Perturbative Boundedness of RG Flow at $\Omega=1$"}
\be
\la_r = -\frac{1}{4\pi^2\theta^2} \frac{\Ga_4(0,0,0,0)}{Z^2} =  1 
+\sum_{n\ge 2} \gamma_n (-\tilde\la)^n 
\ee
where all $\gamma_n$ are finite when the ultraviolet cutoff is removed.

\subsection{The heuristic RG flow} 

\begin{figure}[t]
\centerline{\epsfig{figure=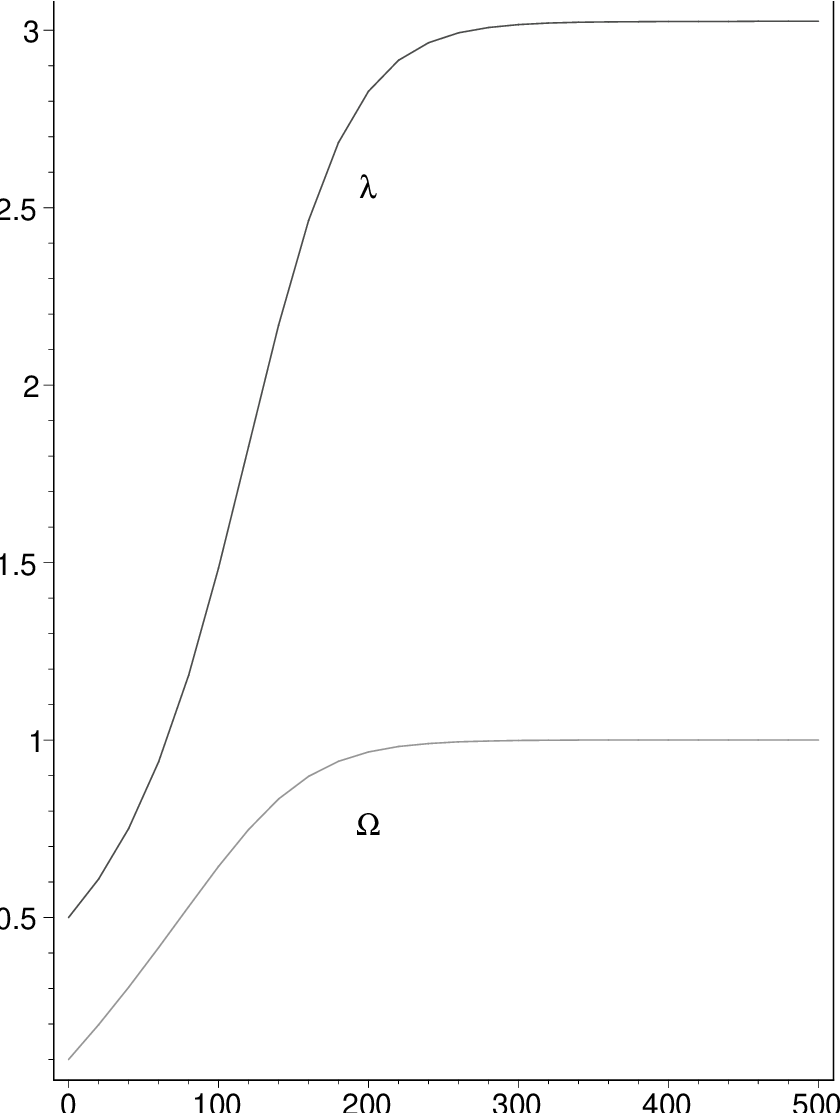 ,width=6cm}}
\caption{Numerical flow for $\la$ and $\Om$}
\label{flow}\end{figure}

If our conjecture is true, not only as a perturbative 
statement at all orders but as a constructive statement,
we should be able to factorize $(1-\Omega)$
in front of the beta function, since the Feynman graphs amplitudes are 
analytic in $\Omega$ near $\Omega=1$. So it is very reasonable that the 
non perturbative RG flow would be:

\bqa
 \frac{d\la_i}{di} &\simeq a (1- \Omega_i) F(\la_i)\ ,\\  
\frac{d\Omega_i}{di} &\simeq b(1- \Omega_i) G(\la_i) \ ,
\eqa
where $F(\la_i) = \la_i^2 + O(\la_i^3)$, $G(\la_i) = \la_i + O(\la_i^2)$
and $a,b\in \mathbb{R}$ are two constants.
The behavior of this system is qualitatively the same as
the simpler system
\bqa  \frac{d\la_i}{di} &\simeq& a (1- \Omega_i) \la_i^2 \ , \\
\frac{d\Omega_i}{di} &\simeq& b (1- \Omega_i) \la_i \ ,
\eqa
whose solution is
\be
\la_i = \la_0 e^{\frac{a}{b} (\Om_i-\Om_0)}\ ,
\ee
with $\Omega_i$ solution of  
\be
b \; i \; \la_0 = \int_{1-\Omega_i}^{1-\Omega_0} e^{\frac{au}{b}} \frac{du}{u}\ ,
\ee
hence going exponentially fast to 1 as $i$ goes to infinity. The
corresponding numerical flow is drawn on Figure \ref{flow}.
\floatplacement{figure}{H}

Of course to establish fully rigorously this picture is beyond the
reach of perturbative theorems and requires a constructive analysis.

\section{Multi-Scale analysis and the Beta function}
\resetequ

The best perturbative expansion is neither the bare expansion, which has no subtractions,
hence no ultraviolet limit, nor the renormalized expansion, which has too many, but the
effective expansion, which has just the necessary ones \cite{R}. Accordingly the best scheme
to compute the evolution of the effective coupling, hence the beta function, is also 
the effective expansion, and it is the one used in this paper. The effective expansion
requires some kind of multiscale analysis. Since at $\Omega=1$ the theory is an exact 
matrix theory, that is integers are conserved along the ribbon sides, hence the faces,
it is convenient to define this multiscale analysis at the level of the variables
associated to each face rather than at the propagator level.

\subsection{Cutoff, Slices} 
Every ribbon graph in this theory has $n$ vertices, $L$ internal lines,
$N$ external legs, $F$ faces, a genus $g$ such that $2-2g=n-L+F$ and 
a number $B$ of broken or "external" faces.  Only graphs with $g=0$, $B=1$ and $N\le 4$
are divergent and enter the RG flow equations.
 
Putting separately the cutoff on the face variables means that the 
theory with ultraviolet cutoff $M^{\rho}$ and infrared cutoff $M^i$, $i<\rho$, is
the sum over all graphs which have all the integers $m$ associated to their
$F-B$ unbroken "internal" faces between $M^{i}$ and $M^{\rho}$: $M^i \le m \le  M^{\rho}$.
Here $m$ means $m_1+m_2$ because Moyal ${\mathbb R}^4$ is made of two symplectic pairs,
hence every face variable is in fact made of two integers $m_1$ and $m_2$.

The part of the theory in the $i$-th slice is the difference between the theory
with infrared cutoff respectively $M^{i+1}$ and $M^i$ so it is 
the sum over all graphs which have all the integers associated to their
$F-B$ unbroken "internal" faces between $M^{i}$ and $ M^{\rho}$,
{\it at least one of them} being exactly between $M^{i}$ and $M^{i+1}$.
 
As well known the effective expansion has only "useful"
renormalization performed. As a consequence it is an expansion
in terms of effective couplings. "Useful" renormalization means
(in our context of cutoffs on faces) that we subtract only
divergent subgraphs with all their internal face variables in higher slices than 
all their external variables.

\subsection{The effective theory} 

Now how does one compute the beta function, namely the flow for the effective vertex 
of the theory?

\paragraph{Mass Renormalization}

To simplify further the effective expansion rules let us remark that 
the mass renormalization is somewhat special since it is the only non-logarithmic.
In any $\phi^4$ theory it is easy to check that 
1PI subgraphs never overlap. Therefore it is not necessary to use 
RG to disentangle "overlapping" mass divergences. Instead 
of using effective masses and useful mass subtractions, it is therefore
convenient to treat the renormalization of mass in the good old pre-Wilsonian
way, namely to use the renormalized mass $\mu$ in the propagator and to mass-renormalize
the perturbation expansion independent of scale attributions (see \cite{R}, end of Chapter II.4). 
Of course this does not work for the coupling constant because there are "overlapping" 
four point divergences, and RG is truly necessary to disentangle them.

So we adopt the rule that in any Feynman graph of the theory any one particle 
irreducible two point subgraph is always subtracted by the corresponding mass 
counterterm, and the subtraction is performed irrespective of the internal and external scales.

\paragraph{Effective Vertices}

Once mass renormalizations have been taken care of, there remains three kinds of
log divergent subtractions: the four point functions,
whose counterterms are of the $\phi^{*4}$ form like the initial Lagrangian interaction,
and the two "two point function second subtractions" which correspond
to the $p^2\phi^2$ and $\tilde x^2\phi^2$ terms of the initial Lagrangian.

If we were to perform the useful subtractions for all these three operators,
we would end up with an effective theory with effective couplings and effective propagators.
But effective propagators are very inconvenient to use (since they are already used for the slicing itself...). 
It is more convenient to remark that at $\Omega = 1$
the $p^2\phi^2$ and $\tilde x^2\phi^2$ terms follow exactly
the same RG trajectory so that their effective coefficients remain equal, at any order (\cite{GWbeta}, last section).  
It corresponds indeed in the language of \cite{GWbeta} to the vanishing of the
$\Omega $ flow at all loops at $\Omega = 1$, which is a simple and direct consequence of 
the exact matrix character of the theory at $\Omega = 1$.
As a consequence we can absorb the effective coefficients $p^2$ and $\tilde x^2$ terms of the initial Lagrangian
into a single "field strength renormalization" which is $Z^{1/2}$, where $Z$ is the propagator renormalization\footnote{This operation in all rigor also changes the parameter $A$ in 
(\ref{propafixed}) into $A/Z$, but this change is inessential for the ultraviolet regime and
our computations below.}.
In this way we keep the propagator coefficients {\it fixed} over all slices.
In other words, and since there are 4 fields per vertex, we can over all RG slices 
use the same fixed propagator (\ref{propafixed}), provided we do not use as effective coupling $\Gamma^4$
but the effective vertex $\lambda_i=[\Gamma^4 /Z^2]_i$.

Now to compute the beta function we have to compute the 
change in this effective vertex when one RG slice $i$ is added in the infrared direction,
and the ultra violet limit $\rho \to \infty $ is performed. 
This change $\lambda_{i}-\lambda_{i+1}$
is therefore the sum over all contributions to this effective vertex 
$\Gamma^4 /Z^2$ which have: 

\begin{itemize}
\item[a)] all their internal faces in slices $j\ge i$, with {\it at least one} exactly in 
the slice $i$,

\item[b)] their (single) external face index taken at the renormalization point, that is at 0 for the
4 point graphs of $\Gamma^4$ and at first derivative at 0 for the 2 point graphs of $Z$, see below,

\item[c)] their mass subtractions performed for all their 1PI two point subgraphs,

\item[d)] their {\it useful} 4 and 2 point log divergent subtractions performed,
that is for all their 2 and 4 point subgraphs {\it whose internal scales are all
strictly bigger than their external scales}.

\item[e)] an effective coupling $\lambda_j$
at each vertex whose maximal index is $j$.
\end{itemize}
The vanishing of the beta function (in the UV regime) means 
that the change $\delta\lambda_i = \lambda_{i}-\lambda_{i+1}$ tends to 0
as $i \to \infty$, and the boundedness of the RG flow is the slightly stronger assumption
that this change not only tends to zero but is summable over $i$. 
Remark that the total flow
$\lambda_{ren}-\lambda_{bare}$ (where $\la_{ren}=\la_0$ and $\la_{bare} 
= \lim_{\rho\to\infty}\la_{\rho}$)
in general is never zero, because 
$\delta\lambda_i $ is not exactly zero, in particular for the first values of $i$. The difference
between the bare and effective vertices is really expressed by the finite $\gamma$ series
in (\ref{mainresult}). 

If the constructive version of this theory
can ever be built as we expect, this $\gamma$ series will hopefully
be shown Borel summable and its Borel sum
should be the non-perturbative finite flow between the bare and the renormalized coupling.

If we inductively know that the beta function vanishes up to order $p$ and study its vanishing
at order $p+1$, then we know that the difference between any effective coupling $\lambda_j$ and the
effective coupling $\lambda_i$ tends uniformly to zero as $i \to \infty$, so that we
can completely forget point e) above and replace all $p+1$ effective vertices $\lambda_j$ 
by a same coupling, eg $\lambda_i$ or $\lambda_{bare}$ itself, 
since the difference will be higher order times finite coefficients. This is what we do below:
we always multiply each contribution of order $p$ by $\lambda^p$ where $\lambda$
is the bare coupling.

Further simplifications are also used below to minimize the amount of actual computations.
In the main identity (\ref{mainresult}), many graphs cancel out completely at the "bare level".
Computing their mass renormalized amplitudes and the useful 
renormalizations for coupling constant and wave function counterterms
would be completely fastidious because these renormalized values always also cancel out if 
the bare values
do! So to simplify the computation we list first all the main
bare (unrenormalized) divergent sums for all possible graphs up to three loops. These graphs
must be planar, which helps make them tractable with some care. 
Then we work out all combinatoric factors and cancel out many bare sums. Only
for the remaining terms (which do not cancel out) do we need
to apply the mass subtractions and the eventual useful log divergent subtractions. 

Up to three loops it happens
that only tadpoles mass subtractions are in fact necessary, because bare amplitudes
with mass "sunshines" insertions such as those of graphs $CS$ and $BS$ cancel out completely
in (\ref{mainresult}). It happens also that only useful
wave-function subtractions of tadpoles are necessary. Of course we suspect that tadpole
subtractions only will not be enough at more than 3 loops.

\section{Basic Sums}
\resetequ

We introduce now systematic notations for the basic integers
sums which are the amplitudes of the two point or four-point Feynman graphs in  this theory. 
These sums without further limitations are divergent, but as explained
in the previous section we will in fact always compute them with one
index (at least) in the $i$-th slice, the others above. Subtractions for the 
mass divergences and for the "useful" subtractions for the log divergences
are always performed even if not explicitly indicated, so that the
real sums which we manipulate are always finite.

This being recalled, the structure of the sums
up to three loops are performed over at most three face integers denoted as $p,q,r$.
They have to be symmetric because our cutoffs
only depend on these face indices (not on the propagators which depend
on the sum of two integers on adjacent faces). 

The upper index in our basic sums is (1), (2) or (3) and recalls the perturbation order.
At order 1 there is only one graph B1 (see Figure \ref{4p12l}). The corresponding sum is:
\be
\cS^{(1)}_1 = \sum_{p} \tfrac{1}{(A+ p)^2} \qquad (B1) \ .
\ee
At order two there are 4 graphs (see Figure \ref{4p12l} again):
\begin{align}
\cS^{(2)}_1 &=[\cS_1^{(1)}]^2= \sum_{p,q} \tfrac{1}{(A+ p)^2}  
\tfrac{1}{(A+ q)^2} \quad (B2)\ , \quad & 
\cS^{(2)}_3 &= \sum_{p,q} \tfrac{1}{(A+ p)^3}   
\tfrac{1}{(A+ q)}   \quad\ \   (TEXT)\ , \nonumber\\
\cS^{(2)}_2 &= \sum_{p,q} \tfrac{1}{(A+ p)^2}  
\tfrac{1}{(A+p+ q)} 
\tfrac{1}{(A+ q)} \quad  \ \ \ (E)\  ,\quad &
 \cS^{(2)}_4 &= \sum_{p,q} \tfrac{1}{(A+ p)^3}  
\tfrac{1}{(A+ p+ q)}   \ \   (TINT)\ . 
\end{align}
At order three there are more graphs (see Figures \ref{4p3l1},\ref{4p3l2}):
\begin{align}
\cS^{(3)}_1 &= \sum_{p,q,r} \tfrac{1}{ (A+p)^2 (A+q)^2 (A+r)^2} = (\Si_1^1)^3 \quad &(B3)  \\ 
\cS^{(3)}_2 &= \sum_{p,q,r} \tfrac{1}{(A+p)^2 (A+q)(A+r)  (A+p+ q)(A+ q+r)}   \quad &(TrE)\\
\cS^{(3)}_3 &= \sum_{p,q,r} \tfrac{1}{(A+p)^2 (A+q)(A+r) (A+p+ q)(A+ p+r)} \quad &(EYB2,\ BEB)\\
\cS^{(3)}_4 &= \sum_{p,q,r} \tfrac{1}{ (A+p)^2(A+q)^2 (A+r)  ( A+q+r)} \quad &(BEY) \\
\cS^{(3)}_5 &= \sum_{p,q,r} \tfrac{1}{(A+p)^2(A+q)^2   (A+p+r)(A+ q+r)} \quad &(EY2)\\
\cS^{(3)}_6 &= \sum_{p,q,r} \tfrac{1}{(A+p)^2(A+q)^2 (A+p+q)(A+ q+r)} \quad &(ETSI)\\
\cS^{(3)}_7 &= \sum_{p,q,r} \tfrac{1}{(A+p)^2(A+q)  (A+p+q)^2(A+ q+r)} \quad &(EITD)\\
\cS^{(3)}_8 &= \sum_{p,q,r} \tfrac{1}{(A+p)^2(A+q)   (A+p+q)^2(A+ p+r)}\quad &(EITG) \\
\cS^{(3)}_9 &= \sum_{p,q,r} \tfrac{1}{(A+p)^2(A+q)^2(A+r)  (A+p+q)} \quad &(ETSE) \\
\cS^{(3)}_{10} &= \sum_{p,q,r} \tfrac{1}{(A+p)^3 (A+q)  (A+p+q)(A+p+r)} \quad &(ETI)\\
\cS^{(3)}_{11} &=\sum_{p,q,r} \tfrac{1}{(A+p)^3 (A+q) (A+r)  (A+p+q)} \quad &(ETE)\\
\cS^{(3)}_{12} &= \sum_{p,q,r} \tfrac{1}{(A+p)^3 (A+q)^2  (A+p+r)} \quad &(B2TI)\\
\cS^{(3)}_{13} &= \sum_{p,q,r} \tfrac{1}{(A+p)^3 (A+q)^2(A+r)   } \quad &(B2TE,\ BT2EE)\\
\cS^{(3)}_{14} &= \sum_{p,q,r} \tfrac{1}{(A+p)^3 (A+q)   (A+p+r)(A+q+r)}\quad &(BS) \\
\cS^{(3)}_{15} &= \sum_{p,q,r} \tfrac{1}{(A+p)^3  (A+p+q)^2 ( A+q+r) } \quad &(BT2II)\\
\cS^{(3)}_{16} &=\sum_{p,q,r} \tfrac{1}{ (A+p)^3 (A+p+q)^2 (A+ p+r) }\quad &(BT2IE) 
\end{align}
\begin{align}
\cS^{(3)}_{17} &= \sum_{p,q,r} \tfrac{1}{ (A+p)^3(A+q)^2 (A+q+r) } \quad &(BT2EIE) \\
\cS^{(3)}_{18} &= \sum_{p,q,r} \tfrac{1}{(A+p)^4 (A+q)  (A+p+r )} \quad &(BEI,\ BIE,\ EBI)\\ 
\cS^{(3)}_{19} &= \sum_{p,q,r} \tfrac{1}{(A+p)^4 (A+p+q )(A+p+r)}\quad &(BII,\ IBI) \\
\cS^{(3)}_{20} &= \sum_{p,q,r} \tfrac{1}{(A+p)^4 (A+q)  (A+r) } \quad &(BEE,\ EBE)
\end{align}
where the names refer to the graphs pictured on the corresponding figures.
We study first $Z$, which involves a smaller number of graphs, and then $\Gamma_4$.

\section{Study of $Z$}
\resetequ

\begin{lemma}  For $\phi^4_4$ and $(\bar\Phi\Phi)^2$ we have
at three loops
\be  
Z = 1 - a   \tilde\lambda  + b  \tilde\lambda^2 - c   \tilde\lambda^3
\ee
where  $\tilde \lambda  = \tfrac{\lambda }{(4 \pi) ^2} $ and  both in the real and complex case
\bqa
a &=& \frac{1}{4}\cS_1^{(1)} \\
b &=&  \frac{1}{16} [\cS^{(2)}_1  +  \cS^{(2)}_2   +2( \cS^{(2)}_3 + 
\cS^{(2)}_4) ]\\
c &= &   \frac{1}{4^3}   
\bigg[(\cS_1^{(3)}+ \cS_3^{(3)}+\cS_4^{(3)}+ \cS_5^{(3)}+\cS_6^{(3)}+\cS_7^{(3)}+\cS_8^{(3)}+
\cS_9^{(3)}) \nonumber\\
&&
+ 2 \big(\cS_2^{(3)}+ \cS_{10}^{(3)} + \cS_{11}^{(3)}
 +\cS_{14}^{(3)} +\cS_{15}^{(3)}
+\cS_{16}^{(3)}+\cS_{17}^{(3)}\big) 
\nonumber\\
&& + 3(\cS_{19}^{(3)}+\cS_{20}^{(3)}) + 4 \cS_{12}^{(3)} +
 6 (\cS_{13}^{(3)}+\cS_{18}^{(3)})\bigg]\ .
\eqa
\end{lemma}
The rest of the section is devoted to the proof.

\subsection{The Self-Energy}

\begin{figure}
\label{tadpoleupdown}
\centerline{\epsfig{figure=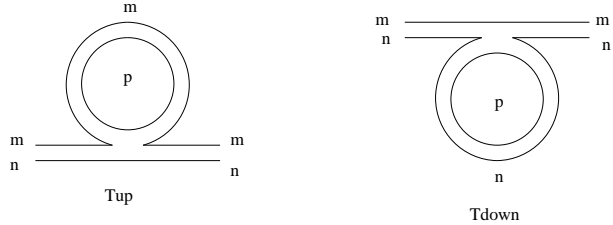,width=8cm}}
\caption{Two Point Graphs at one Loops: the up and down tadpoles}
\label{2p1l}
\end{figure}

\begin{figure}
\label{twopointtwoloops}
\centerline{\epsfig{figure=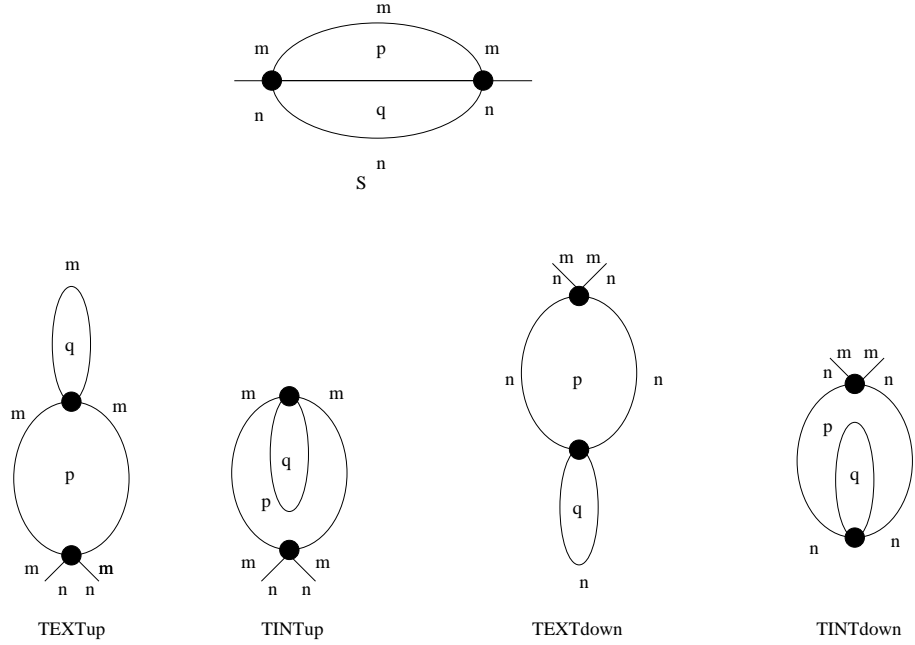,width=12cm}}
\caption{Two Point Graphs at Two Loops}
\label{2p2l}
\end{figure}

\begin{figure}\label{twopointthreeloops}
\centerline{\epsfig{figure=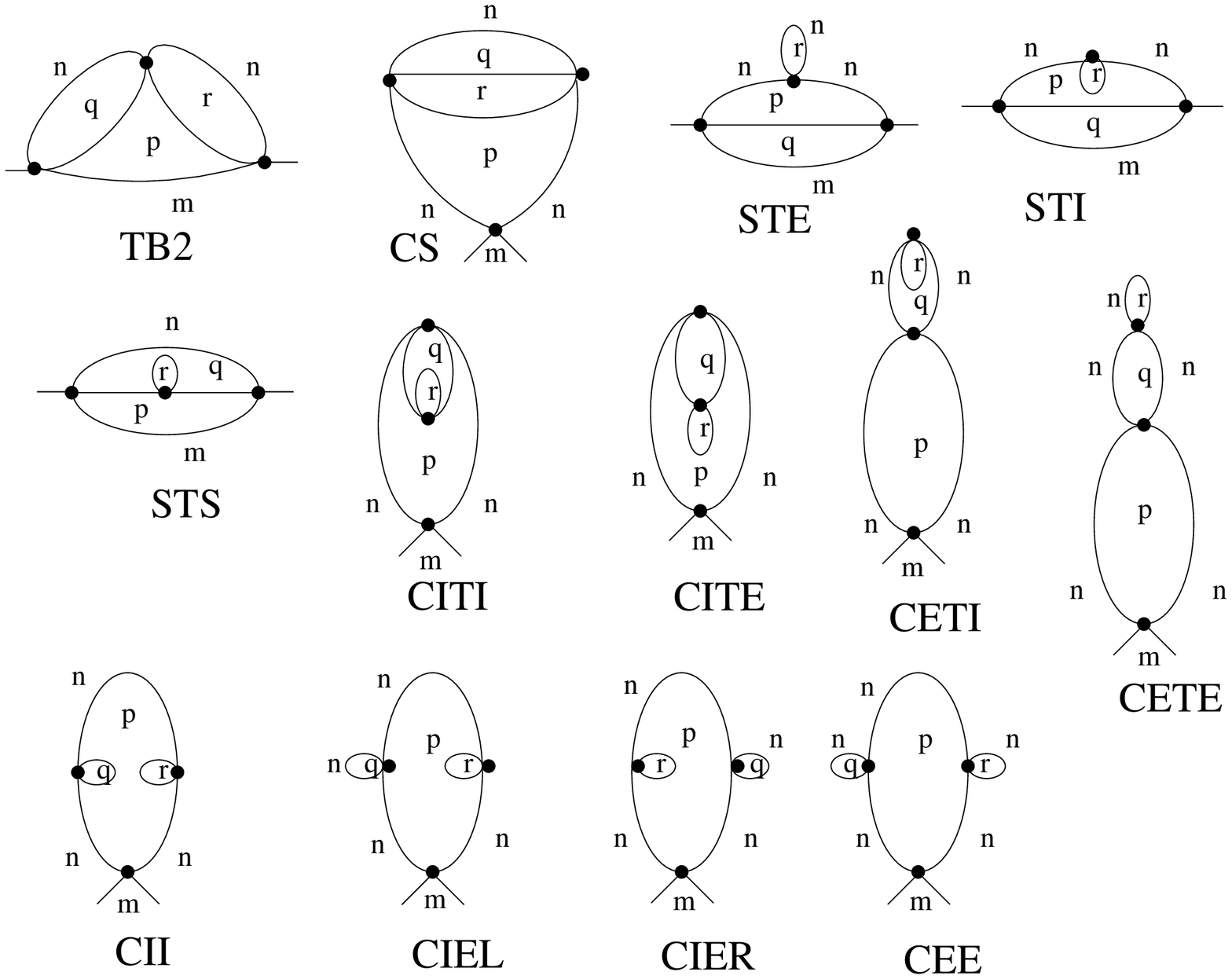,width=11cm}}
\caption{Two Point Graphs at Three Loops}
\label{2p3l}
\end{figure}

\begin{lemma}
The self-energy up to order three, after taking away the factor $ (4\pi)^2 \theta$, is 
\be  
 \tfrac{1}{(4\pi)^2 \theta}\Sigma_{m,n}  = - \tilde\la A_{mn}  + \tilde\la^2 B_{mn} -  
\tilde\la^3 C_{mn}
\ee
where
\begin{align}
 A_{mn} &= \frac{1}{4} \sum_p   [ \underbrace{  \tfrac{1}{A+p+m} }_{Tup }+ 
\underbrace{  m \leftrightarrow n  }_{Tdown } ]  \\
B_{mn}&= \frac{1}{4^2} \sum_{p,q}   [ \underbrace{ \tfrac{1}{(A+p+m)(A+q+n)(A+p+q)}}_{S } +
\underbrace{ ( \tfrac{1}{(A+p+m)^2(A+q+m)}}_{TEXTup }\nonumber \\ &\qquad\qquad \quad +
\underbrace{ \tfrac{1}{(A+p+m)^2(A+p+q)}}_{TINTup}
 + m \leftrightarrow n  ) ] \nonumber 
\end{align}
\begin{align}
C_{mn}&= \frac{1}{4^3} \sum_{p,q,r}   [   
 \underbrace{   \tfrac{1}{(A+p+m)(A+q+n)(A+r+n) (A+p+q)(A+p+r)} }_{TB2 }+ 
\nonumber \\ &+ 
\underbrace{ \tfrac{1}{(A+p+n)^2(A+q+n)(A+q+r) (A+p+r)} }_{CS }
+ \underbrace{ \tfrac{1}{(A+q+m)(A+p+n)^2 (A+p+q)(A+r+n)}}_{STE } +
\nonumber \\ &+  \underbrace{\tfrac{1}{(A+q+m)(A+p+n)^2 (A+p+q)(A+p+r)}}_{STI } 
+   \underbrace{ \tfrac{1}{(A+q+m)(A+p+n)(A+p+q)^2(A+p+r)} }_{STS}  +
\nonumber \\ &+ \underbrace{\tfrac{1}{(A+p+n)^2(A+p+q)^2(A+q+r)}}_{CITI} 
+ \underbrace{ \tfrac{1}{(A+p+n)^2(A+p+q)^2(A+p+r)} }_{CITE} 
+ \underbrace{ \tfrac{1}{(A+p+n)^2(A+q+n)^2(A+q+r)}}_{CETI}  +
\nonumber \\ &
+\underbrace{ \tfrac{1}{(A+p+n)^2(A+q+n)^2(A+r+n)} }_{CETE}  
+ \underbrace{ \tfrac{1}{(A+p+n)^3(A+p+q)(A+p+r)} }_{CII}  
+\underbrace{  \tfrac{2}{(A+p+n)^3(A+q+n)(A+p+r)}}_{CIEL\  + \ CIER}  +
\nonumber \\ &+ \underbrace{ \tfrac{1}{(A+p+n)^3(A+q+n)(A+r+n)} }_{CEE} 
+ m \leftrightarrow n  ]
\end{align}
and the corresponding graphs are listed in Figures \ref{2p1l},\ref{2p2l} and \ref{2p3l}.

\end{lemma}

\paragraph{Proof}

\noindent At {\bf one loop} we have only one vertex and one line, therefore, after
taking out the factor  $ (4\pi)^2 \theta$ we have
\be
 -\tilde\lambda (4 \pi^2 \theta^2) (4\pi)^2 \tfrac{1}{[(4\pi)^2 \theta]^2} 
\sum_{G} \tfrac{K_G}{4} (S_G^{(1)})_{mn} 
=  -\frac{\tilde\lambda}{4} \sum_{G} \tfrac{K_G}{4} (S_G^{(1)})_{mn} 
\ee
where $G$ are the graphs at one loop (there are two of them, see Figure \ref{2p1l}), 
$(S_G^{(1)})_{mn} $ are the
corresponding amplitudes (listed above) and $K_G$ is a combinatorial factor.
The combinatorial factors for $T\ up$ and $T\ down$ are the same
\be
K_G^r  = 4\ \Rightarrow \frac{K^r_G}{4}=1\ .
\ee
In the complex case we have instead of $K_G^r/4$ the term $K_G^c/2$ and $K_G^c=2$.
So in both cases we have
\be
  A_{mn}=\frac{1}{4} \sum_{G}  (S_G^{(1)})_{mn}  \ .
\ee

\vskip 0.5cm

\noindent At {\bf two loops} we have two vertices and three lines, therefore, after
taking out the factor  $ (4\pi)^2 \theta$ we have
\be
\tfrac{1}{2!}\tilde\lambda^2  (4 \pi^2 \theta^2)^2 (4\pi)^4 
\tfrac{1}{[(4\pi)^2 \theta]^4} \sum_{G} \tfrac{K_G}{4^2} (S_G^{(2)})_{mn} 
=  \tfrac{\tilde\lambda^2}{4^2} \tfrac{1}{2}\sum_{G} \tfrac{K_G}{4^2} (S_G^{(2)})_{mn} 
\ee
where $G$ are the graphs at two loops of Figure \ref{2p2l}, $(S_G^{(2)})_{mn} $ are the
corresponding amplitudes (listed above) and $K_G$ is again a combinatorial factor.
As before, in the complex case instead of $K_G^r/4$ we have  $K_G^c/2$.
The factors (in the real case) are
\be
K_S^r= 2\cdot 4^2\ ,\quad K_{T2EXTup}^r =  2\cdot 4^2\ , \quad   K_{T2INTup}^r  =  
2\cdot 4^2 \ .
\ee
The same factors hold for the graphs with $m$ and $n$ exchanged.
In the complex case we have 
\be
K_S^r= 2\cdot 2^2\ ,\quad K_{T2EXTup}^r =  2\cdot 2^2\ , \quad   K_{T2INTup}^r  =  
2\cdot 2^2 \ .
\ee
so in both cases we have
\be
B_{mn} =  \frac{1}{4^2} \sum_{G}  (S_G)^{(2)}_{mn} \ .
\ee
\vskip 0.5cm

\noindent At {\bf three loops} we have three vertices and five lines, therefore, after
taking out the factor  $ (4\pi)^2 \theta$ we have
\be
-\frac{1}{3!}\tilde\lambda^3  (4 \pi^2 \theta^2)^3 (4\pi)^6 
\frac{1}{[(4\pi)^2 \theta]^6} \sum_{G} \frac{K_G}{4^3} (S_G^{(3)})_{mn} 
=  -\frac{\tilde\lambda^3}{4^3} \frac{1}{3!}\sum_{G} \frac{K_G}{4^3} (S_G^{(3)})_{mn} 
\ee
where $G$ are the graphs at three loops (Figure \ref{2p3l}), $(S_G^{(3)})_{mn} $ are the
corresponding amplitudes (listed above) and $K_G$ is again a combinatorial factor.
As before, in the complex case instead of $K_G^r/4$ we have  $K_G^c/2$.
The factors (in the real case and complex case) are
\be
K_G^r= 3!\ 4^3\ ,\qquad K_G^r= 3!\ 2^3 \qquad \forall G \ ,
\ee
therefore in both cases
\be
C_{mn} =  \tfrac{1}{4^3} \sum_{G}  (S_G)^{(3)}_{mn} \ .
\ee
This completes the proof of the lemma.

\subsection{Proof of lemma 1}
Now from the definition of $Z$, to prove lemma 1 we need to compute

\begin{align}
a=& -\left . \partial_{m_1} A_{mn}\right |_{m=n=0} =  
-\left .\partial_{n_1} A_{mn}\right |_{m=n=0} \ ,
\nonumber\\
b=& -\left .\partial_{m_1}B_{mn}\right |_{m=n=0} =  -\left.\partial_{n_1}B_{mn}\right |_{m=n=0}\ ,
\nonumber\\
c=& -\left .\partial_{m_1}C_{mn}\right |_{m=n=0} =  -\left .\partial_{n_1} C_{mn}\right |_{m=n=0} \ .
\end{align}
From the definitions of $A_{mn}$, $B_{mn}$ and $C_{mn}$ we get 
\be
a =   \frac{1}{4} \sum_p   [ \underbrace{  \tfrac{1}{(A+p)^2} }_{Tup }+  
\underbrace{ 0 }_{Tdown }]
\ = \  \tfrac{1}{4} \cS_1^{(1)}  \ .
\ee
\begin{align}
b =& \frac{1}{4^2} \sum_{p,q,r} [ 
 \underbrace{ \tfrac{1}{(A+p)^2(A+q)(A+p+q)}}_{S }  
 + \underbrace{ \left ( 2 \tfrac{1}{(A+p)^3(A+q)}+\tfrac{1}{(A+p)^2(A+q)^2} \right ) }_{TEXTup }+
\underbrace{2 \tfrac{1}{(A+p)^3(A+p+q)}}_{TINTup}
 + 0 ) ]    \nonumber\\
= &  
 \frac{1}{16} [\cS^{(2)}_1 + \cS^{(2)}_2 + 2 (\cS^{(2)}_3 + 
\cS^{(2)}_4)]\ .
\end{align}

\begin{align}
c &= \frac{1}{4^3} \sum_{p,q,r}   \bigg[   
 \underbrace{ \cS_3^{(3)}  +2  \cS_2^{(3)} }_{TB2 +  m \leftrightarrow n }+ 
\underbrace{ 2 \cS_{14}^{(3)} + \cS_{5}^{(3)} }_{CS  +  m \leftrightarrow n }
+ \underbrace{ \cS_{9}^{(3)} + 2\cS_{11}^{(3)} + \cS_{4}^{(3)} }_{STE +  m \leftrightarrow n } +
\nonumber \\ &+  \underbrace{ 
 \cS_{6}^{(3)} + 2 \cS_{10}^{(3)}  }_{STI +  m \leftrightarrow n } 
+   \underbrace{ \cS_{8}^{(3)}+  \cS_{7}^{(3)}}_{STS +  m \leftrightarrow n}  +
 \underbrace{2 \cS_{15}^{(3)} }_{CITI+  m \leftrightarrow n} 
+ \underbrace{ 2 \cS_{16}^{(3)} }_{CITE+  m \leftrightarrow n} 
+ \underbrace{ 2\cS_{17}^{(3)} + 2  \cS_{12}^{(3)}}_{CETI+  m \leftrightarrow n}  +
\nonumber \\ &
+\underbrace{ 4\cS_{13}^{(3)} + \cS_{1}^{(3)} }_{CETE+  m \leftrightarrow n}  
+ \underbrace{
3 \cS_{19}^{(3)} }_{CII+  m \leftrightarrow n}  
+ \underbrace{  
2(3 \cS_{18}^{(3)} +  \cS_{12}^{(3)})}_{CIEL\  + \ CIER+  m \leftrightarrow n}  +
 \underbrace{
3  \cS_{20}^{(3)} +  2\cS_{13}^{(3)}  }_{CEE+  m \leftrightarrow n}     \bigg] \ ,
\end{align}
so that putting these together
\begin{align} 
c &=  \frac{1}{4^3}   \big[(\cS^{(3)}_1+ \cS^{(3)}_3+\cS^{(3)}_4
+ \cS^{(3)}_5+\cS^{(3)}_6+\cS^{(3)}_7+\cS^{(3)}_8+\cS^{(3)}_9)  + 2 (\cS^{(3)}_2+ \cS^{(3)}_{10}
\nonumber\\
&  + \cS^{(3)}_{11}+\cS^{(3)}_{14} +\cS^{(3)}_{15}+\cS^{(3)}_{16}+\cS^{(3)}_{17}) 
+ 3(\cS^{(3)}_{19}+\cS^{(3)}_{20})+ 4\cS^{(3)}_{12} + 6 (\cS^{(3)}_{13}+\cS^{(3)}_{18})\big]\ .
\end{align}

\section{Study of $\Ga_4$}
\resetequ

\begin{lemma}  For both  $\phi^4_4$ and $(\bar\Phi\Phi)^2$ we have
at three loops, before performing the mass renormalization
\be  
\Gamma_4 (0,0,0,0)  = - \lambda (4 \pi ^2 \theta^2 ) 
[1 -a' \tilde\lambda  +b' \tilde\lambda^2  - c'\tilde\lambda^3  ]
\ee
where $\tilde \lambda  = \tfrac{\lambda }{(4 \pi) ^2} $ and

\bqa
a' &=& \frac{1}{2} \cS_1^{(1)}\\
b' &=& \frac{1}{8} [\cS^{(2)}_1+ 2 (\cS^{(2)}_2+\cS^{(2)}_3+\cS^{(2)}_4)]\\
c' &=& \frac{1}{32}\bigg[(\cS^{(3)}_1 + \cS^{(3)}_5) + 
2 (\cS^{(3)}_4  + \cS^{(3)}_6  + \cS^{(3)}_7   + \cS^{(3)}_8 + \cS^{(3)}_9   
\nonumber \\
&& \quad \ +  \cS^{(3)}_{14}   + \cS^{(3)}_{15}   + \cS^{(3)}_{16}   + \cS^{(3)}_{17}  ) +
3(\cS^{(3)}_3   + \cS^{(3)}_{19}  +  \cS^{(3)}_{20} )
\nonumber \\
&& \quad \ +  4 
(\cS^{(3)}_{4}  +\cS^{(3)}_{10}  + \cS^{(3)}_{11}   + \cS^{(3)}_{12} )  + 
6  (\cS^{(3)}_{13} + \cS^{(3)}_{18})\bigg]
\eqa
and the corresponding graphs are drawn in Figure \ref{4p12l}, \ref{4p3l1} and \ref{4p3l2}.

\end{lemma}
The rest of this section is devoted to the proof.

\subsection{One Loop}
\begin{figure}
\centerline{\epsfig{figure=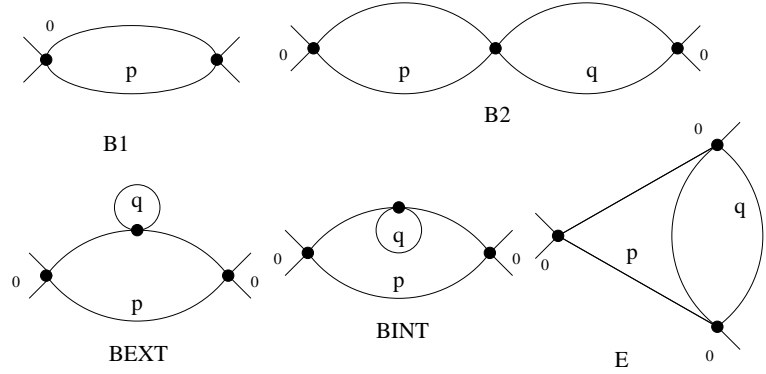,width=10cm}}
\caption{Four Point Graphs at one and two Loops}
\label{4p12l}\end{figure}

At one loop there is only one graph contributing to $\Ga_4$, the bubble
$B1$ in Figure \ref{4p12l}.

This graph has two vertices and two lines, plus a factor $1/2!$ from
the exponential. After taking out the factors 
$ - \lambda (4 \pi ^2 \theta^2 (- \tilde\lambda)$ we have, in the real
case
\be
a'\  =\  \frac{1}{8}  \left ( \frac{K_r(B)}{4^2}\right )  
\sum_{p\in\N^2} \tfrac{1}{(A+p)^2}   
\ =\  \frac{1}{8}  \left (\frac{K_r(B)}{4^2}\right )  \cS_1^{(1)} \ ,
\ee
where  $K_r(B)=4^3$ is the combinatoric factor counting the number
of times this graph appears. In the complex case we get the same
expression, except that instead of $K_r(B)/4^2$ we have
$K_c(B)/2^2$ and the combinatoric factor is now $K_c(B)=2^4$.
\be 
\frac{K_r(B)}{4^2} = \frac{K_c(B)}{2^2} = 4 \ .
\ee 
The result is 
\be
a' = \frac{1}{2} \sum_p \tfrac{1}{(A+p)^2}  = \frac{1}{2}  \cS^{(1)}_1  = 2a \ .
\ee

\subsection{Two Loops}
At two loops there are four graphs  $B2$ (double bubble), $E$ (eye), 
$BEXT$, $BINT$ (see Figure \ref{4p12l}). So in the real case
\be  
b' = \frac{1}{3!\ 4^2} \sum_{G}\frac{K_r(G)}{4^3}  S_G(p,q) \ ,   
\ee
where
\begin{align}
&S_{B2}=  \sum_{p,q}  \tfrac{1}{(A+p)^2}   \tfrac{1}{(A+q)^2}  =
\Sigma^2_1  = (\cS_1^{(1)})^2 \ ,  
&&S_{E} =\sum_{p,q} \tfrac{1}{(A+p)^2}   \tfrac{1}{A+q}   \tfrac{1}{(A+p+q)}  
 = \cS^{(2)}_2 \ , \nonumber \\
&S_{BEXT} = \sum_{p,q}\tfrac{1}{(A+p)^3}   \tfrac{1}{A+q}   = \cS^{(2)}_3  \  , 
&&S_{BINT} =\sum_{p,q} \tfrac{1}{(A+p)^3}   \tfrac{1}{A+p+q}  = \cS^{(2)}_4  \ .
\end{align}
In the complex case we have the same expression with $K_c(G)/2^3$ instead of
$K_r(G)/4^3 $.
The combinatorial coefficients in the real case are
\be
K_r(E) = K_r(BEXT) =K_r(BINT) = 3! \cdot  4^3\cdot 4\ ,\qquad 
K_r(B2) = 3! \cdot 4^3  \cdot 2 \ ,
\ee
and in the complex one are
\be
K_c(E) = K_c(BEXT) =K_c(BINT) = 3! \cdot  2^3\cdot 4\ ,\qquad 
K_c(B2) = 3! \cdot 2^3  \cdot 2 \ ,
\ee
so
\be
\tfrac{K_r(E)}{4^3} = \tfrac{K_c(E)}{2^3} = 
\tfrac{K_r(BEXT)}{4^3} = \tfrac{K_c(BEXT)}{2^3} = 
\tfrac{K_r(BINT)}{4^3} = \tfrac{K_c(BINT)}{2^3} = 3! \cdot 4 \ , 
\nonumber\ee
\be
\tfrac{K_r(B2)}{4^3} = \tfrac{K_c(B2)}{2^3} = 3! \cdot 2 \ .
\ee
and 
\be 
b' =  \frac{1}{8}[\cS^{(2)}_1  + 2 (\cS^{(2)}_2   + \cS^{(2)}_3 + \cS^{(2)}_4) ] \ .
\ee

\subsection{Three Loops}
\begin{figure}
\centerline{\epsfig{figure=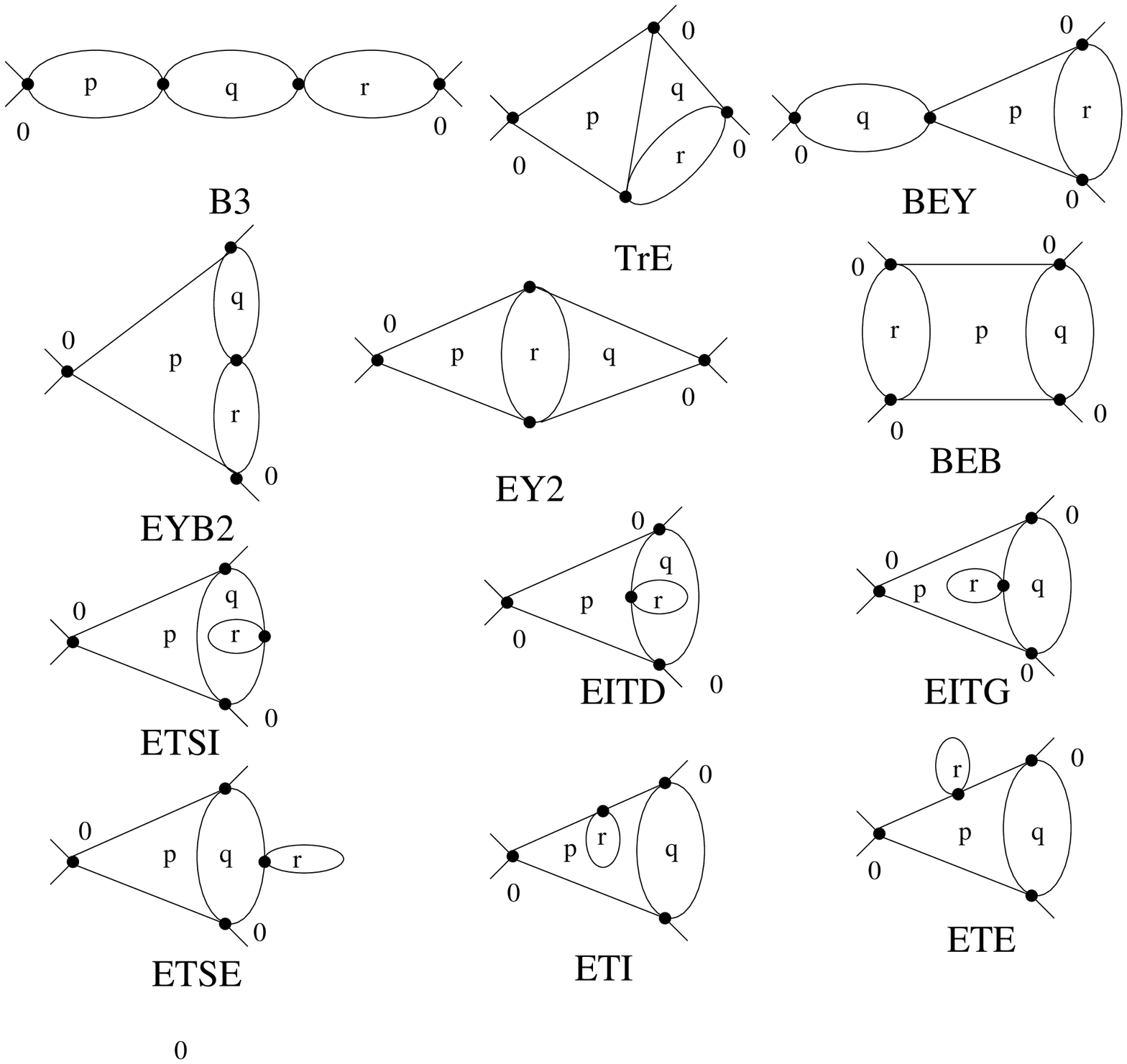,width=10cm}}
\caption{Four Point Graphs at Three Loops, Part I}
\label{4p3l1}\end{figure}

\begin{figure}
\centerline{\epsfig{figure=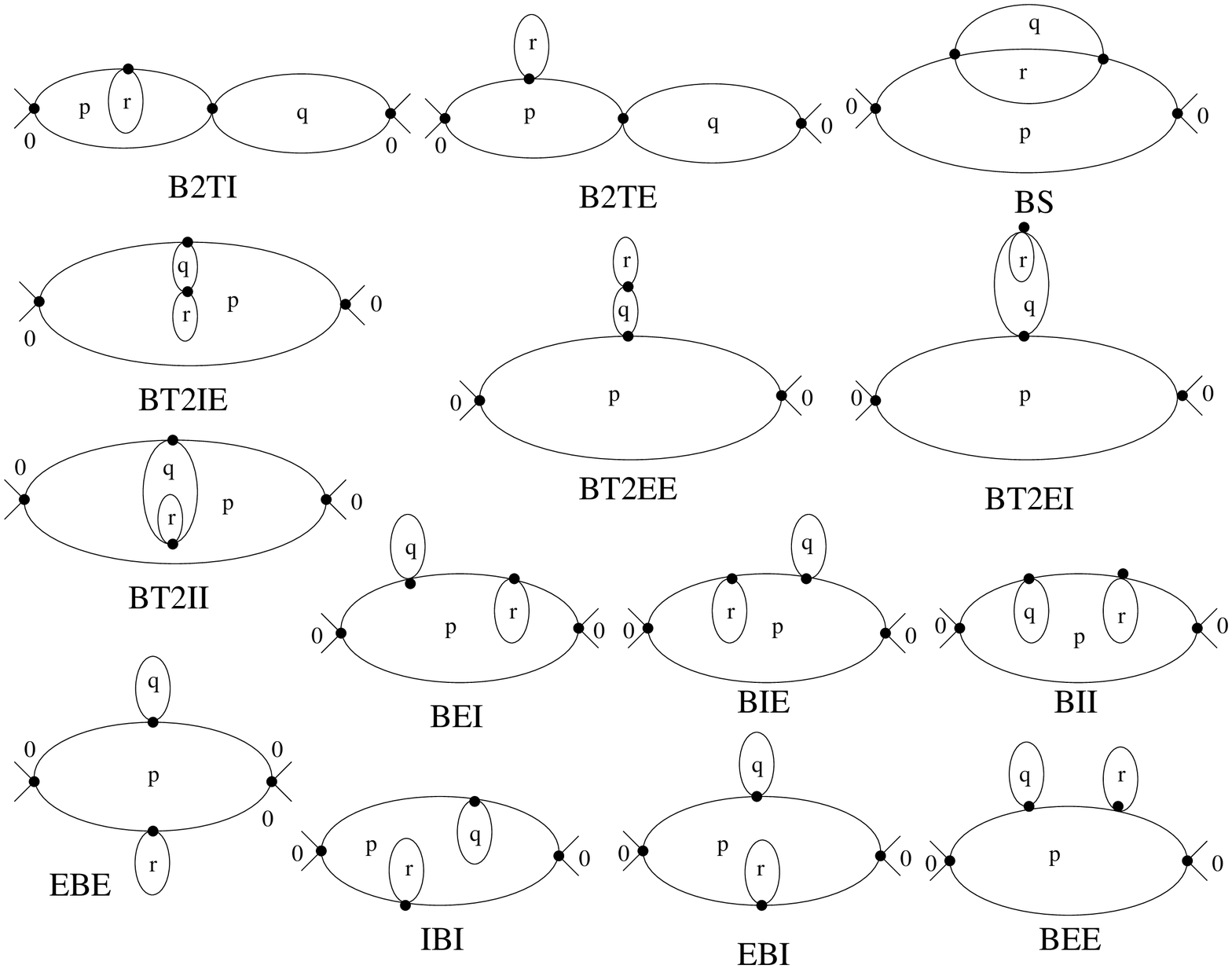,width=10cm}}
\caption{Four Point Graphs at Three Loops, Part II}
\label{4p3l2}\end{figure}

At three loops the 26 graphs
contributing to $\Gamma_4$ are drawn with their code names in 
Figures \ref{4p3l1} and \ref{4p3l2} .

After eliminating $-\lambda 4\pi^2 \theta^2  $
from $(-\lambda 4\pi^2 \theta^2)^4   
\tfrac{1}{4!} \tfrac{1}{[(4\pi)^2)\theta]^6}  \tfrac{1}{4^4}$
and a $-(\tilde\lambda)^3 $ the value of $c'$ is
\bqa c'= \frac{1}{4!}  \frac{1}{4^3} 
\sum_{i=1}^{20} \frac{K_i}{4^{4}} \cS^{(3)}_i \ .
\eqa

For the real case we  have 
\begin{align}
K_1 &=  4! 4^4 \cdot 2 \qquad (B3) & 
K_2 &=  4! 4^4 \cdot 4\cdot2 \quad (TrE) \\
K_3 &=  4! 4^4 \cdot 6 \qquad (4 EYB2+ 2  BEB)  &
K_4 &=  4! 4^4 \cdot 4 \qquad (BEY)\nonumber \\
K_5 &=  4! 4^4 \cdot 2 \qquad (EY2) & 
K_6 &=  4! 4^4 \cdot 4 \qquad (ETSI)  \nonumber \\
K_7 &=  4! 4^4 \cdot 4 \qquad (EITD) &
K_8 &= 4! 4^4 \cdot 4  \qquad (EITG)\nonumber  \\ 
K_9 &=  4! 4^4 \cdot 4 \qquad  (ETSE) &
K_{10} &= 4! 4^4\cdot 8 \qquad (ETI)&\nonumber \\
K_{11} &= 4! 4^4\cdot 8 \qquad (ETE) &
K_{12} &= 4! 4^4\cdot 8 \qquad (B2TI)  \nonumber \\
K_{13} &=  4! 4^4\cdot 12 \quad \ (B2TE + BT2E) &
K_{14} &=  4! 4^4\cdot 4 \qquad (BS)  \nonumber  \\
K_{15} &= 4! 4^4\cdot 4 \qquad (BT2II) &
K_{16} &= 4! 4^4\cdot 4 \qquad  (BT2IE) \nonumber\\ 
K_{17} &= 4! 4^4\cdot 4 \qquad  (BT2EI) &  
K_{18} &=  4! 4^4\cdot 12 \quad  \ (BEI+ BIE+ EBI)\nonumber\\
K_{19} &= 4! 4^4\cdot 6\qquad  (BII+IBI) & 
K_{20} &= 4! 4^4\cdot 6\qquad   (BEE + EBE)\nonumber 
\end{align}
In the complex case the coefficients are the same, except the
$4^4$ factor becomes a $2^4$ one (at each vertex we have two choices instead of
four to contract a particular field), so both in the real and complex case we have 
\begin{align} 
c' &= 
  \frac{1}{32}[(\cS^{(3)}_1 + \cS^{(3)}_5) + 2 (\cS^{(3)}_4 + 
\cS^{(3)}_6 + \cS^{(3)}_7  + \cS^{(3)}_8+ \cS^{(3)}_9  
+ \cS^{(3)}_{14}  + \cS^{(3)}_{15}  + \cS^{(3)}_{16}  + \cS^{(3)}_{17} ) 
\nonumber \\
&+  
3(\cS^{(3)}_3  + \cS^{(3)}_{19} +  \cS^{(3)}_{20})+ 4 (\cS^{(3)}_2 +
\cS^{(3)}_{10}  + 
\cS^{(3)}_{11}  + \cS^{(3)}_{12} )  + 
6  (\cS^{(3)}_{13} + \cS^{(3)}_{18})].
\end{align}

\section{Beta function and Proof of the Theorem}
\resetequ

For the beta function the important combination is
\bqa  
- \frac{1}{ \lambda (4\pi^2\theta^2)}\frac{\Gamma_4}{Z^2} &=& 
\frac{1- a' \tilde \lambda + 
b'\tilde \lambda^2 - c'\tilde \lambda^3}{[1- a \tilde \lambda + b \tilde \lambda^2 -  
c \tilde \lambda^3]^2} 
\\  \nonumber
&=& [1 - \gamma_1  \tilde\lambda +\gamma_2  \tilde\lambda^2 - \gamma_3  \tilde\lambda^3 + O(\la^4)]
\eqa
and we need to prove that each $\gamma$ is finite. In fact it turns out that 
since $a'=2a$, $\gamma_1=0$
so multiplying by $Z^2$ and expanding out, the equations to prove are:
\bqa
a'=2a\ ; \ b'= 2b + a^2 + \gamma_2\ ; \ c' = 2c +2ab + 2a\gamma_2 + \gamma_3 \ .
\eqa

\subsection{One and Two loops}
Let us prove the two first equations.
For $\Omega=1$, $a=a'/2$, hence the first equation $a'=2a$  holds  and
\bqa  
b' - 2 b + 3 a^2 - 2 aa'    &=& b' - 2 b -\tfrac{1}{4} {a'}^2   =
\tfrac{1}{8}[\cS^{(2)}_1  + 2 (\cS^{(2)}_2   + \cS^{(2)}_3 + \cS^{(2)}_4) ]
\nonumber\\
&& - \frac{1}{8} [\cS^{(2)}_1 + \cS^{(2)}_2 + 2 (\cS^{(2)}_3 + \cS^{(2)}_4)]
 - \frac{1}{16} (\cS_1^{(1)})^2 \ \nonumber\\
&=& \ \frac{1}{16}[2\cS^{(2)}_2 -  \cS^{(2)}_1 ] \ .
\eqa
But
\bqa  2\cS^{(2)}_2 -  \cS^{(2)}_1 &=& \sum_{p,q}  \tfrac{1}{(A+p)^2} 
\tfrac{1}{(A+q)^2}  \tfrac{1}{A+p+q} 
[A + (q-p)] \nonumber\\
&=& A \sum_{p,q}  \tfrac{1}{(A+p)^2} \tfrac{1}{(A+q)^2}  \tfrac{1}{A+p+q} ,
\eqa
so that the second equation holds with  
\be  \gamma_2 =  A \sum_{p,q}  \tfrac{1}{(A+p)^2} \tfrac{1}{(A+q)^2}  \tfrac{1}{A+p+q} \ > 0\ .
\ee

\subsection{Three Loops}

The key to check our theorem at 3 loops (taking into account $2a=a'$) is to check that 
$ c' -a'b -2 c - a'\gamma_2 = -\gamma_3$ is finite.

From previous
\be
a'b = \tfrac{1}{32}\cS_1^{(1)} (\cS^{(2)}_1 +\cS^{(2)}_2
+2(\cS^{(2)}_3 +\cS^{(2)}_4) )= \tfrac{1}{32}(\cS^{(3)}_1 +\cS^{(3)}_4+
2(\cS^{(3)}_{12} +\cS^{(3)}_{13}) )
\ee
hence
\bqa
c'  -a' b -2 c &=& - \cS^{(3)}_1 + \cS^{(3)}_6+\cS^{(3)}_7+\cS^{(3)}_8+\cS^{(3)}_9 \nonumber\\ 
&+& 2 (\cS^{(3)}_2 + \cS^{(3)}_3 + \cS^{(3)}_{10} +  \cS^{(3)}_{11} - 
\cS^{(3)}_{12} - \cS^{(3)}_{13})\ .
\eqa
Now it is convenient to rewrite $a' \gamma_2$ as 
\be
\frac{1}{32}\sum_{pqr} \frac{1}{(A+r)^2}  [ 2 \cS^{(2)}_2 - \cS^{(2)}_1]
= \frac{1}{32} [2 \cS^{(3)}_4 - \cS^{(3)}_1]
\ee 
to get 
\be
c'  -a' b -2 c  - a'\gamma_2 =  \cS^{(3)}_6+\cS^{(3)}_7+\cS^{(3)}_8+\cS^{(3)}_9  + 
2 (\cS^{(3)}_2 + \cS^{(3)}_3 + \cS^{(3)}_{10} +  \cS^{(3)}_{11} - 
\cS^{(3)}_{4} - \cS^{(3)}_{12} - \cS^{(3)}_{13})\ .
\ee

From now on let us apply the necessary mass renormalizations. The renormalized sums are
\be
[\cS^{(3)}_9]_{ren}  = [\cS^{(3)}_{11}]_{ren} = [\cS^{(3)}_{13}]_{ren}  = 0  \ ,
\ee
\begin{align}
[\cS^{(3)}_{6}]_{ren} &= -\cS^{(3)}_{2} + \cR_6\ ,\quad & 
[\cS^{(3)}_{7}]_{ren} &= -\Delta\cS^{3}_{7}  + \cR_7\ ,\nonumber\\
[\cS^{(3)}_{8}]_{ren} &= -\Delta\cS^{3}_{8}  + \cR_8\ ,\quad &
[\cS^{(3)}_{10}]_{ren} &= -\cS^{3}_{3}  + \cR_{10}\ , \nonumber\\
[\cS^{(3)}_{12}]_{ren} &= -\cS^{3}_{4}  + \cR_{12}\ , 
\end{align}
where
\begin{align}
\Delta\cS^{3}_7 =& \sum_{pqr}  \tfrac{1}{(A+p)^2 (A+p+q)^2(A+r)(A+q+r)}\ ,
&   \Delta\cS^{3}_8 =& \sum_{pqr}  \tfrac{1}{(A+p)(A+q) (A+p+q)^2(A+r)(A+p+r)}\ ,\nonumber\\
\cR_6 =& \sum_{pqr} \tfrac{A}{(A+p)^2 (A+q)^2 (A+p+q)(A+r) (A+q+r)} \ ,
  &   \cR_7 =& \sum_{pqr} \tfrac{A}{(A+p)^2 (A+q) (A+p+q)^2(A+r)(A+q+r)}\ , 
\nonumber\\   \cR_8 =& \sum_{pqr} \tfrac{A}{(A+p)^2 (A+q) (A+p+q)^2(A+r)(A+p+r)}\ , 
  &    \cR_{10}=&\sum_{pqr}\tfrac{A}{(A+p)^3 (A+q) (A+p+q)(A+r)(A+p+r)} \ ,
\nonumber\\   \cR_{12}=&\sum_{pqr}\tfrac{A}{(A+p)^3 (A+q)^2(A+r)(A+p+r)}\ . & & 
\end{align}

Hence  after symmetrization
\begin{align}
  \gamma_3 =& -\cS^{(3)}_{2} + \cR_6  -\Delta\cS^{3}_{7}+ \cR_7-\Delta\cS^{3}_{8} + \cR_8
\nonumber\\ &
+2 (\cS^{(3)}_2 + \cS^{(3)}_3 - 
\cS^{(3)}_{4} -\cS^{(3)}_{3}+ \cR_{10}+ \cS^{(3)}_{4}- \cR_{12})\nonumber\\
=&+\cS^{(3)}_{2}-\Delta\cS^{3}_{7}-\Delta\cS^{3}_{8} + \cR_6  + \cR_7 + \cR_8
+ 2 (\cR_{10}- \cR_{12})\ .
\end{align}
We have
\be 
\cS^{(3)}_{2}-\Delta\cS^{3}_{7}-\Delta\cS^{3}_{8} \ =\  - \sum_{pqr}  
\tfrac{A}{(A+p)^2 (A+q)(A+p+q)^2(A+r)(A+q+r)}
\ =\  - \cR_7\ ,
\ee
so that
\be \gamma_3= 2 (\cR_{10}- \cR_{12}) + \cR_6  + \cR_8 \ .
\ee
Now
\be
\cR_{10}- \cR_{12} = - \cR_6 + \cR'\ , 
\qquad
\cR_{8}-\cR_6 = -\cR_7 + \cR''\ ,
\ee
\begin{align}
\cR'=&\sum_{pqr}\tfrac{A^2}{(A+p)^3 (A+q)^2 (A+p+q) (A+r)(A+p+r)}\ ,\nonumber\\
\cR''=& \sum_{pqr}\tfrac{A^2}{(A+p)^2 (A+q)^2 (A+p+q)^2(A+r) (A+p+r)}\ ,
\end{align}
so that
\be \gamma_3= -\cR_7 + (2 \cR' + \cR'')\ .
\ee
This is still log divergent when $r>p$ and $r>q$, 
 because we have not yet  performed the necessary wave-function renormalizations.
Practically this consists in subtracting the factor $1/(A+r)^2$ to $1/(A+r)(A+p+r)$
(or $1/(A+r)(A+q+r)$)
 when $r>p,q$. The resulting expressions are finite.

\subsubsection*{Acknowledgment}
We thank R. Gurau, J. Magnen, F. Vignes-Tourneret and R. Wulkenhaar for useful discussions.

\end{document}